# Visual Selective Attention System to Intervene User Attention in Sharing COVID-19 Misinformation

Zaid Amin[1] , Nazlena Mohamad Ali[2*] , Alan F. Smeaton[3]

Institute of IR4.0 (IIR4.0), Universiti Kebangsaan Malaysia, Malaysia[1, 2]
Faculty of Informatics Engineering, Universitas Bina Darma, Palembang, Indonesia[1]
INSIGHT: Centre for Data Analytics, Dublin City University, Dublin 9, Ireland[3]

*Abstract*—Information sharing on social media must be accompanied by attentive behavior so that in a distorted digital environment, users are not rushed and distracted in deciding to share information. The spread of misinformation, especially those related to the COVID-19, can divide and create negative effects of falsehood in society. Individuals can also cause feelings of fear, health anxiety, and confusion in the treatment COVID-19. Although much research has focused on understanding human judgment from a psychological underline, few have addressed the essential issue in the screening phase of what technology can interfere amidst users' attention in sharing information. This research aims to intervene in the user's attention with a visual selective attention approach. This study uses a quantitative method through studies 1 and 2 with pre-and post-intervention experiments. In study 1, we intervened in user decisions and attention by stimulating ten information and misinformation using the Visual Selective Attention System (VSAS) tool. In Study 2, we identified associations of user tendencies in evaluating information using the Implicit Association Test (IAT). The significant results showed that the user's attention and decision behavior improved after using the VSAS. The IAT results show a change in the association of user exposure, where after the intervention using VSAS, users tend not to share misinformation about COVID-19. The results are expected to be the basis for developing social media applications to combat the negative impact of the infodemic COVID-19 misinformation.

## I. INTRODUCTION

The disruption of major changes in the digital environment makes social media applications an "omnipresence" in human life. This change also affects the overload of information on the internet. According to Statista [1], there are currently 4.2 billion active users of social media applications, and this number will remain to grow. As a result of this diverse digital environment, users relatively get distracted, primarily when they receive and share information on social media [2]. The behavioral factor of human attention has long been a key factor in science and research that focuses on Human-Computer Interaction (HCI). The study conducted by [3] states that the attention factor's role is essential when users share information on social media. In line with that, recent studies conducted by Gabielkov et al. [4] stated that about 59% of users in Twitter share information without even reading the content first (in other words, in a hurry and without attentive behavior to share the information they have just received).

Several technological innovations have been developed, one developed by Facebook, which relies on an algorithm to detect false information. However, the approach that relies on robot-based applications needs to be re-examined by carrying out additional "hybrid" integration, specifically considering the psychological factors of human decisions. This is in line with research by [5], where they found that the spread of false information was carried out by humans more than bot-based applications. This shows that the spread of false information on social media requires a lot of collaborative studies and research that can understand human decision factors.

The spread of misinformation, especially regarding COVID-19, can have a multidimensional negative impact on society. These negative impacts include false information about treatment, belief in certain drugs and medical treatments, economic incentive motives for pharmaceutical companies, and polarization of mental exposure [6] and [7]. Furthermore, the destructive impact of spreading COVID-19 misinformation can affect the mitigation process's pace in handling democracy and economic recovery in a country, for example, in expediting the implementation of COVID-19 vaccination to the public.

Several studies from [8] - [10] state that this "visual selective attention" technique can influence user decisions when facing a task. Therefore, in this study, we build a tool with a visual selective attention technique to intervene the user's attention when deciding to share information on social media. In the context of this study, users deal with ten information they receive using the Visual Selective Attention System (VSAS). In each pre-and post-intervention task, the VSAS will stimulate the user with an interface using "spotlight" and "zoom-lens" design techniques [11], where these techniques are a sub-theory of visual selective attention, which has been known to play an important role in decision-making.

In this study, we aim to 1) intervene in user attention using VSAS when they will share misinformation about COVID-19, 2) measure the effectiveness of VSAS in intervening users, 3) measure user association and evaluation when they will share misinformation about COVID-19 using Implicit Association Test (IAT). This paper is divided into six sections. Section 1 contains the introduction. Section 2 contains relevant and related studies. Section 3 contains the methods used in study 1 and study 2. Section 4 contains the results of analysis and experiments. Section 5 contains discussions and research limitations, and section 6 contains conclusions.

*Corresponding Author.

## II. BACKGROUND WORK

### A. User Attention Factor

The study in [2] shows that one of the key factors underlying why users share misinformation on social media is that users' attention is distracted, and when users do not think critically about the information they receive, they are likely to share the information.

The attention factor is one of the fundamental psychological factors in humans when interacting with their environment. The attention factor is a set of cognitive processes that make a person able to process a set of information in limited conditions either because the capacities of environment or the cognitive state s/he has [12]. From another perspective, the attention factor is related to a person's level of awareness and focus when receiving and confirming information [13].

A study by [14] found that the role of the attention factor is vital in explaining the phenomenon of how users behave in online media. This attention behavior relates to when a user reads tweets, surfs any websites, and accesses e-mails. The study by [15] states that social media designers need to maximize users' level of attention and awareness when accessing the information on social media. For example, design properties that can stimulate user visibility when using social media applications can be shape or pattern components with contrasting color strengths.

### B. The influence of Selective Attention on user Decisions

An example of a concrete concept of selective attention technique is when we get a pop-up message from sending an email. The design of the pop-up message with a "quick display" design can distract us and influence our decision to open the incoming email. This visual selective attention technique was also described by [9] when designing different interfaces in the form of multi-display patterns and locations. This technique can increase the user's attention when searching for information. As for the health and medical aspects, Lopes and Ramos [16] found that selective attention, integrated into the health application interface, showed significant results, particularly increasing user attention in understanding health literacy.

### C. The influence of Social Influences and Epistemic Belief in Sharing Information

According to Chen et al. [17], when sharing information on social media, users are very easily influenced by the social influence factor. For example, if users receive information from related or emotionally close people, they are more likely to trust and re-share the information. This is in line with the study conducted by [18], who found that the effect of self-actualization only appeared when the user focused on close friends (focused on bonding social relationships). Based on this, we then included the narrative stimuli of social influence in pre- and post-intervention sessions using VSAS.

Another psychological factor that can trigger the possibility of sharing information without attention is the epistemic belief factor. A study conducted by Chua and Banerjee [19] stated that epistemic belief factors could influence user decisions to share health rumors online. Users with confidence and experience with certain drugs or medical treatments tend to be easily influenced when processing false health information and are more likely to share it again.

The theoretical basis and the role of the epistemic belief factor align with the Implicit Association Test (IAT) measurement method, where the IAT measures the level of the user's association tendency or exposure to a concept in a person. In this paper context, the user is expected to evaluate the ten-information provided in pre-and post-intervention using VSAS.

## III. METHODS

### A. VSAS Tool

In the VSAS experiment, we conducted pre-and post-intervention sessions for two weeks. Thirty-eight participants joined in this experimental session. Participants (n=38) consisted of 11 females and 27 males with an average age of 20 (SD = 1.11), and all were students. Participants were compensated $5 for their time. Each experiment section took 1.5 to 2 hours. In the first week, we used VSAS without any design intervention. In the second week, the intervention was carried out using visual selective attention by adding a label design to the ten-information provided to participants. This label design was generated based on credible fact-checking sources. The same ten items of information in weeks 1 and 2 were given to participants, while the content of the information relates to the political, sensational, and sensitive context [20].

We built and designed this VSAS using JAVA programming through the Android Studio 4.1.1 application. The storage media for each response from our participants uses the services of the Firebase Real-Time Database. This VSAS instrument concept is built and designed for mobile-based applications. The layout of the VSAS instrument wireframe design can be seen in Figures 1 and 2.

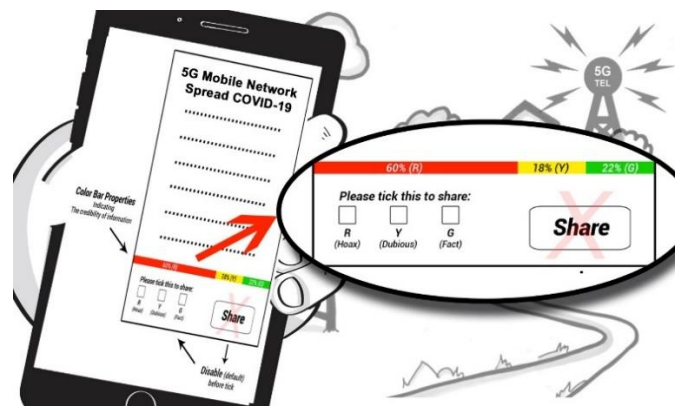

Fig. 1. Wireframe Design using Spotlight Technique.

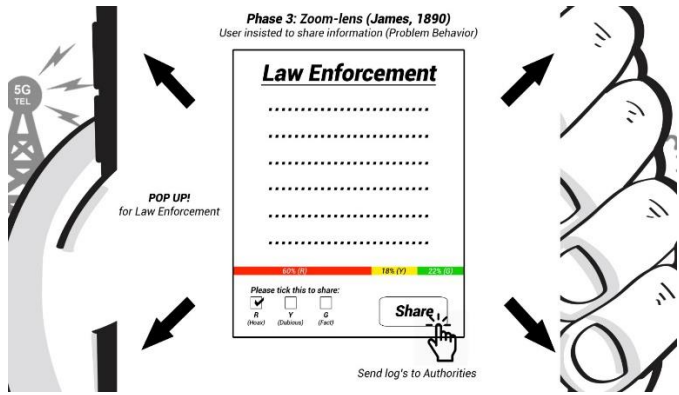

Fig. 2. Wireframe Design using Zoom-Lens Technique.

*B. Procedures*

In the first week's session (pre-intervention), each participant registered in advance to communicate with the administrator. After 38 participants were registered and had an account in VSAS, they sent their respective profile and demographic data. Then, administrators sent participants through individual chats of ten information contexts, of which five items were misinformation, and the other five were fact information. Each participant was asked to respond to the chat from the administrator by replying using a Likert-based scale between 1 - 5. Participants will be asked and narrated at each delivery of context information, representing the social influence factor like "if you got this information from your family, would you share it again." To answer these questions, participants replied with the Likert scale between 1 - 5, where "1" indicates strongly disagrees with sharing information, "2" disagrees with sharing information, "3" is neutral, "4" agrees to share information, and "5" for strongly agrees to share information.

After participants responded to ten information contexts in the pre-intervention session, we stored the respondent's data in the Firebase Real-Time Database for tabulation and analysis. After the experiment using VSAS is carried out, the participants will start the Implicit Association Test (IAT) session through the Pavlovia website. The IAT content is related to how participants evaluate "Misinformation vs. Fact Information or Positive vs. Negative Words." In this IAT session (see Figure 3), participants will quickly determine the information according to their respective perceptions. The process of associating factual information and misinformation is combined with the participant's ability to determine "Positive vs. Negative Words."

In the second week session (post-intervention), participants will be sent the same ten contexts of information as in session 1. The difference is that the concept of visual selective attention through spotlight and zoom-lens design techniques has been applied. The attention of participants will be intervening by focusing on the design of the spotlight color bar properties, where red color stimuli indicate the label "hoax," yellow is "dubious," and green is "fact" (see Figure 4).

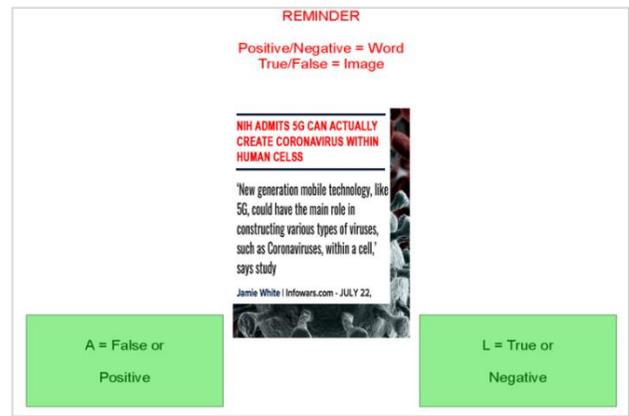

Fig. 3. Example of IAT Test Selection.

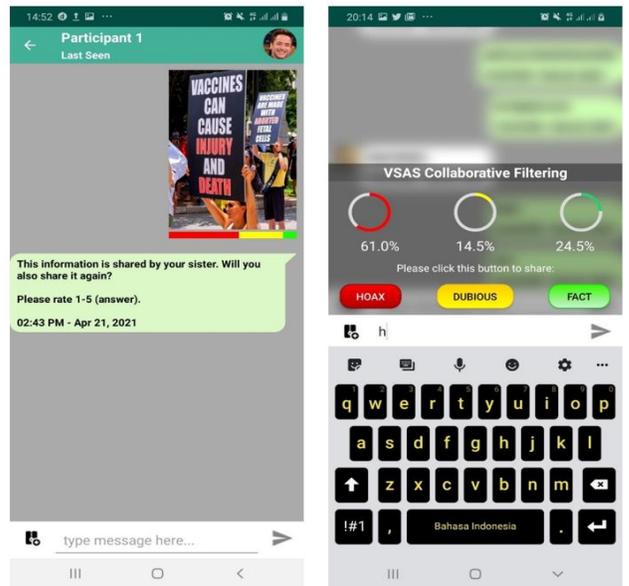

Fig. 4. Intervention Process on VSAS.

The labeling method for each of the ten information contexts is carried out based on fact-checking sources. In this intervention session, participants will also be asked to be involved in providing collaborative corrections. Next, at the last intervention stage, participants will be intervened with pop-up warning notifications in the form of law enforcement. This pop-up law enforcement increases each participant's attention with the "zoom-lens" animation technique. After the post-intervention was completed, participants would start the IAT with the same as the first-week session.

IV. RESULTS

We summarized using descriptive statistical analysis to briefly examine the pre-and post-intervention results (see Figure 5). After conducting the second-week session (post-intervention), only 23 participants completed the entire VSAS and IAT experimental process, and 15 participants experienced errors. Therefore, we could not analyze the data responses.

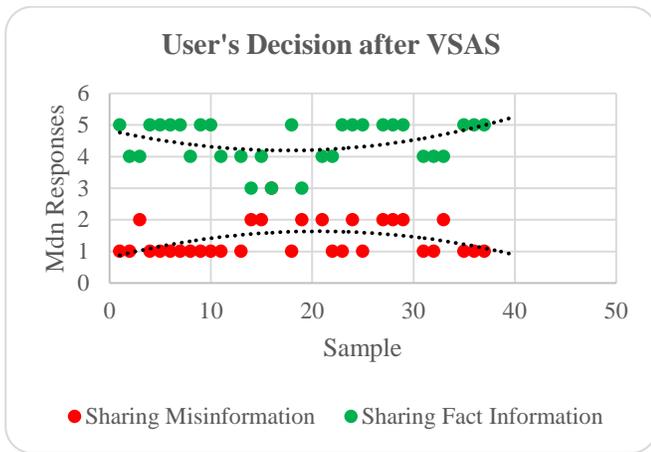

Fig. 5. User Response during Post-Intervention.

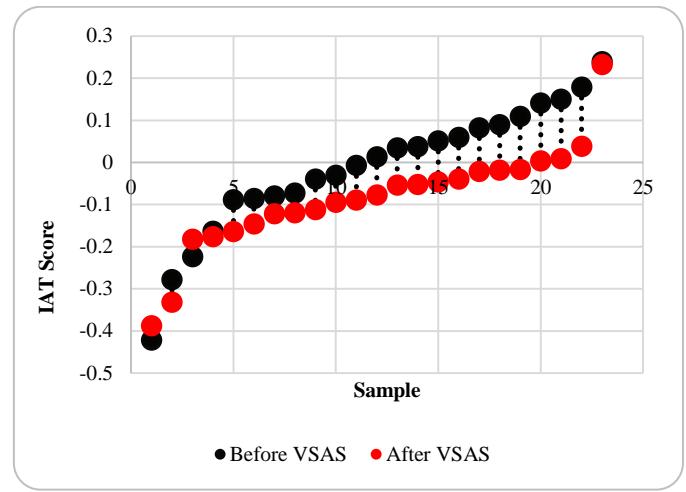

Fig. 6. User Evaluation during Post-Intervention.

TABLE I. DISTRIBUTION OF IMPLICIT SCORES AFTER VSAS

| D-score | Category | n | Percent |
|---|---|---|---|
| -2 to -0.65 | Strong negative | 0 | 0.0 |
| -0.65 to -0.36 | Moderate negative | 1 | 2.6 |
| -0.35 to -0.15 | Slight negative | 4 | 10.5 |
| -0.15 to 0.15 | Neutral/ No Preference | 17 | 44.7 |
| 0.15 to 0.36 | Slight positive | 1 | 2.6 |
| 0.36 to 0.65 | Moderate positive | 0 | 0.0 |
| 0.65 to 2 | Strong positive | 0 | 0.0 |
|  | Missing | 15 | 39.5 |
| **Total** |  | **38** | **100.0** |

The significant results in 23 participants in post-intervention showed a behavior change, whereas 15 participants increased their attention and chose to confidently respond with an answer of "1" for each context of the misinformation sent. The results in Figure 6 describe the successions of the VSAS experiments, which is 65.2% of participants strongly disagreed with sharing the misinformation provided.

This result asserts that VSAS has succeeded in intervening in users' attention when they decide to share information. A total of 7 participants, or 30.4%, chose answer "2" where they did not agree to reshare the misinformation they received. A total of 22 participants significantly chose answers in the Likert range of 1-2, especially about the context of misinformation on implementing COVID-19 vaccination. In the descriptive analysis of the post-intervention results using VSAS, we can conclude that most of the participants' decision tendencies are Mdn=1, IQR=1. The results of IQR=1 indicate that the distribution of the participants' median responses at post-intervention also shows a linear result, which has less variability about its median.

Finally, to ensure the success of changing responses and evaluating participants on the IAT, we analyzed the IAT calculations using a D-score. This D-score (see Table 1) is similar to the Cohen's d effect measurement, ranging from -2.00 to 2.00. In detail, the frequency distribution of the D-score category shows as 17 participants in the "Neutral/No Preference" D-Score category, 4 participants in the "Slight negative" D-Score category, and 1 participant in the "Slight positive" D-Score category. Based on these results (see Figure 7), we can state that the use of VSAS can validly improve participant evaluation in the context of measuring the concept of "Misinformation vs. Fact Information or Positive vs. Negative Words."

In the last IAT analysis step, we calculated the correlation between the IAT Score and the MdnScore (Median Score). We then also calculated the percentage weights for each question's score, calculated the variance and the ranking (see Figure 7). The correlation between IAT Score and MdnScore (Median) in sharing misinformation showed a significant moderate positive correlation of $p = +0.3344$.

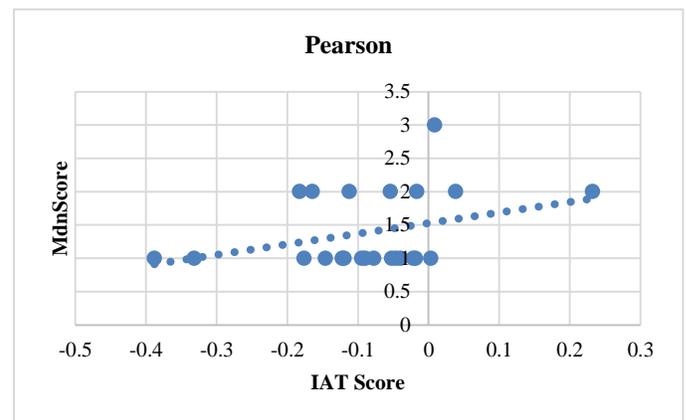

Fig. 7. The Correlation between IAT Score and Median in Sharing Misinformation.

## V. DISCUSSION AND LIMITATIONS

The previous two studies have determined relevant attention-based design, including Implicit Association Test principles, later implemented on the Visual Selective Attention System (VSAS) while evaluating information about COVID-19. These findings are useful to inform researchers/designers on design requirements that should consider in developing an attention-based design that significantly influences user decision on sharing COVID-19 misinformation. This study has also stated that visual techniques have a more optimal role in influencing user behavior and decisions on a task. For this reason, we recommend that this visual technique could be the preference when building a technology related to user behavior in sharing information on social media. Several techniques or designs in building attention-based interfaces need further exploration and adaptation. The stimulant technique using sound or video media and the multiscreen technique might be preferred in future research. This study only carried out VSAS interactions in the distribution of 10 (ten) information contexts and social influence narratives within the scope of "individual chats." In future research, we suggest that the process of sending ten contexts of information and social influence narratives can be carried out within the scope of a "group chat." The point is to determine how different user responses are between in/out-group with the same interests or beliefs. In designing and building VSAS applications, we have limitations in obtaining secondary data in the form of datasets from social media platforms. In future studies, this dataset can be used as a reference and comparison to understand the diffusion process and information dissemination patterns in social media. Knowing this real-world situation will enhance an entire perspective in developing VSAS applications.

## VI. Conclusion and Future Work

Based on the results of this study, we conclude that VSAS can increase attentive behavior when deciding to share misinformation about COVID-19. The future development of VSAS requires a more extensive study of understanding other psychological factors that influence user attention when deciding to share information. The results of this study can be the basis for developing social media applications that can be used in a wider domain, not only in the context of the COVID-19 issue, but also in the context of other domains such as security issues, handling disaster mitigation, and others, especially in communication management in handling crisis.

In future research, it is necessary to have categories of participants with different and diverse demographic backgrounds. This aims to enrich the experimental results and develop the features in VSAS, especially in selecting intervention techniques to be carried out. Research collaboration is needed to understand the essence of other key psychological concepts related to attentional behavioral factors in sharing information on social media. Meanwhile, to measure in detail, future research needs to consider how to measure the *"attention span"* aspect and its relationship with a person's critical thinking ability. This is important so that further research can know the intervention's effectiveness precisely and then will be able to answer the challenge of how quickly the intervention process can be carried out in the context of user interaction when sharing information. The research results reported in this study ultimately clearly show that the VSAS system has succeeded in changing user behavior in deciding whether to share information, in line with the IAT results, which show significant changes in user tendencies while evaluating information about COVID-19.

ACKNOWLEDGMENT

This work was supported by the Universiti Kebangsaan Malaysia research grant under Grant GPK-4IR-2020-019.

## AUTHOR'S PROFILE

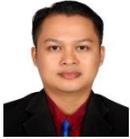

ZAID AMIN is currently pursuing the Ph.D. degree with the Institute of IR4.0, Universiti Kebangsaan Malaysia. He is currently a Lecturer with the Faculty of Computer Science, Universitas Bina Darma, Indonesia. He is very enthusiastic about human-computer interaction. His research interests include interaction design, UI/UX, persuasive technology, and social computing

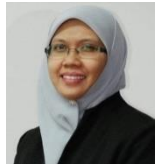

NAZLENA MOHAMAD ALI received the Ph.D. degree in human-computer interaction from Dublin City University, Ireland, in 2009. She is currently an Associate Professor and a Senior Research Fellow with the Institute of IR4.0, Universiti Kebangsaan Malaysia. Her research interests include interaction design, UI/UX, persuasive technology, digital games, and user engagement.

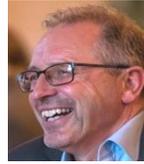

ALAN F. SMEATON (Fellow, IEEE) is currently a Professor of computing and the Former Director of the Insight-Centre for Data Analytics, Dublin City University. His research interests include human memory, why we forget some things and not others, and how we can use technology like search systems, to compensate for when we do forget. He was the winner of the Royal Irish Academy Gold Medal for Engineering Sciences, in 2015. He is the Chair of ACM SIGMM and an IEEE Fellow.